\documentstyle[pra,aps,multicol,epsf]{revtex}
\begin{document}
\draft
\title{Spin Anisotropy and Quantum Hall Effect in the Kagome Lattice
\protect\\
- Chiral Spin State based on a Ferromagnet-}
\author{Kenya Ohgushi, Shuichi Murakami,  and Naoto Nagaosa}
\address{Department of Applied Physics, University of Tokyo,
Bunkyo-ku, Tokyo 113-8656, Japan}
\date{\today}
\maketitle
\begin{abstract}
A ferromagnet with spin anisotropies on the 2D Kagome lattice is
theoretically studied. This is a typical example of the 
flat-band ferromagnet. The Berry phase induced by the tilting
of the spins opens the band gap and quantized Hall conductance
$\sigma_{xy}=\pm e^2/h$ is realized {\it without} external magnetic
field. This is the most realistic chiral spin state based on the
ferromagnetism. We also discuss the implication of
our results to anomalous Hall effect observed in the metallic
pyrochlore ferromagnets $R_2$Mo$_2$O$_7$ $(R=$Nd, Sm, Gd$)$.

\end{abstract}
\pacs{ 75.10.Lp, 75.30.Gw, 72.10.-d, 72.15.Eb}

\begin{multicols}{2}
\narrowtext
The spin Berry phase plays an important role in the quantum transport 
of strongly correlated electronic systems.
Consider an electron  hopping  from site $i$ to $j$ coupled 
to a spin at each site with Hund's coupling 
$J_H$ \cite{andersonhasegawa}. When $J_H$ is strong enough the spin of
the hopping electron
is forced to align parallel to $\vec S_i$ and $\vec S_j$ at each site,
with the spin wave function being $| \chi_i>$ and $| \chi_j>$, respectively.
The spin wavefunction is explicitly given by
\begin{equation}
\vert \chi_i \rangle = {}^t \biggl[ e^{i b_i} \cos{ {\theta_i} \over 2},
                e^{i(b_i+ \phi_i)} \sin{ {\theta_i} \over 2} 
\biggr],
\end{equation}
where we have introduced the polar coordinates as
$\langle \chi_i \vert \vec S_i \vert \chi_i \rangle =\frac{1}{2} 
(\sin \theta_i \cos \phi_i, \sin \theta_i \sin\phi_i,
\cos \theta_i )$. The overall phase $b_i$  corresponds to the gauge 
degrees of freedom and does not appear in  physical quantities.
Therefore, the effective transfer integral $t_{ij}$ is given by
\cite{andersonhasegawa}
\begin{eqnarray}
t_{ij} &=&  t \langle \chi_i | \chi_j  \rangle 
\nonumber \\
&=& t e^{i ( - b_i + b_j)} \biggl[ \cos{ {\theta_i} \over 2} 
\cos{ {\theta_j} \over 2} +
e^{i(-\phi_i+ \phi_j)} \sin{ {\theta_i} \over 2} \sin{ {\theta_j} \over 2} 
\biggr]
\nonumber \\
&=& t e^{i a_{ij}} \cos{ {\theta_{ij}} \over 2}, 
\end{eqnarray}
where $\theta_{ij}$ is the angle between the two spins $\vec S_i$
and  $\vec S_j$.
The phase $a_{ij}$ is the vector potential generated by the spin,
and corresponds to the Berry phase felt by the hopping electron.
Let us consider an electron hopping along a loop $1 \to 2 \to 3 \to 1$.
The total phase that the electron obtains is the 
gauge flux by $a_{ij}$, which  corresponds to the solid angle
subtended by the three spins $\vec S_i$ ($i = 1,2,3)$.
This is called the spin chirality and is one of the key concept in the 
physics of strongly correlated electronic systems
\cite{kalmeyer,baskaran,laughlin,wwz,leenagaosa}. 
One can easily see that the spin chirality is absent for collinear
spin alignment, and the spin chirality has been intensively 
discussed in the context of quantum spin liquid  where
the spins and hence $a_{ij}$ fluctuate quantum mechanically
\cite{kalmeyer,baskaran,laughlin,wwz,leenagaosa}. 
This $a_{ij}$ is the leading actor in the gauge theory of 
strongly correlated electronic systems \cite{baskaran,leenagaosa}. 

Among them the proposal of chiral spin liquid with broken time-reversal
symmetry in a triangular  lattice \cite{kalmeyer}, and later the anyon
superconductivity \cite{laughlin}
attracted great interests at the early stage of the high-$T_c$ research.
Wen et al. \cite{wwz} constructed a mean field theory for a chiral spin liquid
on a square lattice. They start from the $\pi$-flux state,
and break the time-reversal symmetry by introducing the
next-nearest-neighbor hopping with a phase.
However, it turned out to be rather difficult to find  physical 
realization of the (chiral) spin liquid in real materials, even in
frustrated lattices.

The spin Berry phase has been discussed also in the context of anomalous
Hall effect (AHE) in manganites \cite{ye}. It is proposed that the
spin-orbit interaction $H'$ leads to the coupling between the 
magnetization $M$ and the spin chirality, i.e., the gauge flux, $b$
as expressed by $ H' = \lambda b M$.
At finite temperature $T$, Skyrmions are thermally excited
and the balance between the positive and negative chiralities  
is broken by $H'$ to give rise to a  finite average $\langle b \rangle$.
This $\langle b \rangle$ gives an additional ``magnetic field'' and hence
leads to the anomalous Hall effect proportional to the 
coupling constant $\lambda$ and Skyrmion density 
$\sim e^{- \Delta/T}$, where $\Delta$ is the excitation energy of
the Skyrmion. 
This mechanism is the novel one coming from the Berry phase of the spins 
compared with the conventional skew-scattering mechanism  \cite{ahe,kl,kondo}.
However it shares a feature with these conventional theories,
namely the AHE vanishes in the zero-temperature limit, which
is the case in the conventional ferromagnetic metals experimentally \cite{ahe}.
It is related to the fact that the spin chirality is zero in the ground state.
 However, it is noted that a recent work proposes the
staggered flux state as the ground state of the double exchange model on
a cubic lattice with doping \cite{yamanaka}.
  
On the other hand, recent transport experiments on  ferromagnetic  
pyrochlores $R_2$Mo$_2$O$_7$ ($R=$Nd, Sm, Gd) revealed that the  
AHE increases as $T$ is lowered and approaches to the 
saturated value \cite{taguchi,katsufuji}. This behavior is qualitatively
different from the conventional one \cite{ahe}.
One clue to explain this anomalous feature is that the pyrochlore 
structure has geometrical frustration \cite{review}.
It consists of corner-sharing tetrahedrons and the antiferromagnetic
interactions between nearest neighbor spins are frustrated. It was recently
pointed out that even the ferromagnetic interaction  is frustrated,
if the spin easy-axis  points to the center of the tetrahedron
\cite{harris,moessner}. In this case, because the spin configuration becomes
non-collinear, we expect the spin chirality appears and
affects the quantum transport of electrons, especially the transverse
conductivity $\sigma_{xy}$. However, it is a nontrivial
issue whether the spin chirality really contributes to 
 $\sigma_{xy}$ when the spins form a periodic
structure.

Motivated by these pyrochlore compounds, we study in this letter the
two-dimensional Kagome lattice, which is the cross section of 
the pyrochlore lattice perpendicular to the (1,1,1)-direction \cite{review}.
We show that the chiral spin state is realized
in an {\it ordered} spin system on the Kagome lattice,
when the spin anisotropy is 
introduced. When the Fermi energy is in the gap, the system shows 
quantized Hall effect {\it without} external magnetic field.
Implications of our results to these pyrochlore compounds
are also discussed especially on the AHE
which does not vanish at low temperatures.
\begin{figure}
\begin{center}
\vspace{0mm}
\hspace{0mm}
\epsfxsize=7cm
\epsfbox{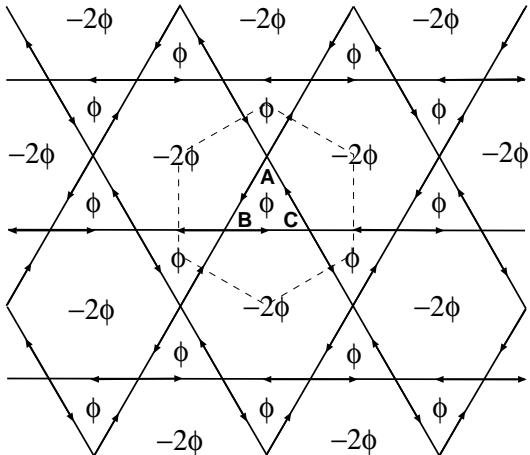}
\caption{Kagome lattice. The dotted line represents the Wigner-
Seitz unit cell, which contains three independent sites (A,B,C).
It is assumed that each site has different spin anisotropy axis.
The arrows on bonds mean the sign of the phases of the transfer
integral $t_{ij}$.}
\label{kagome}
\end{center}
\end{figure}

\begin{figure}
\begin{center}
\vspace{0mm}
\hspace{0mm}
\epsfxsize=7cm
\epsfbox{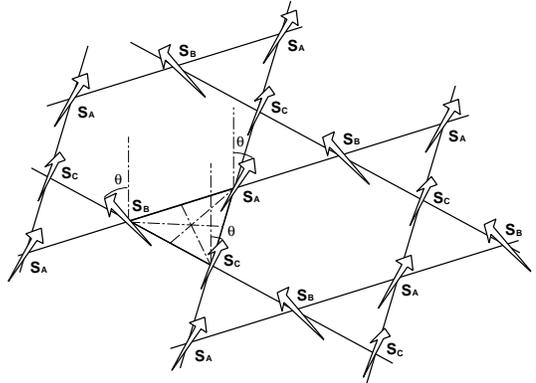}
\caption{One possible spin configulations. Each spin is tilted due to
the spin anisotropy to the center of triangles by $\theta$ from z-axis.}
\label{spin}
\end{center}
\end{figure}

We consider the tight-binding model on the Kagome lattice 
shown in Fig.\ref{kagome}.
The unit cell contains three sites (A,B,C), and we put three spins on
each site. These spins are assumed to be ferromagnetically coupled with
each other, and are fully polarized in the ground state. Since three sites on a
triangle are crystallographically independent,
the spin anisotropy on each sites are expected to be 
different and produce the spin chirality. As an example one possible spin
configurations are presented in Fig.\ref{spin}.
The tight-binding model for the electrons strongly Hund-coupled to these
spins is given by
\begin{equation}
H =  \sum_{i j} t_{ij} C^\dagger_i C_j
\label{eqn:tight}
\end{equation}
with $t_{ij}$ being given by eq.(2). 
The phase of $t_{ij}$ is 
the same for all the nearest neighbor pairs with the direction shown by
the arrows in Fig.\ref{kagome}. We set the flux  originated from
the spin chirality
per triangle as $\phi$, which satisfy
$e^{i \phi} =e^{i(a_{AB}+a_{BC}+a_{CA})}$. Especially in the case of
Fig.\ref{spin}, $\phi=\pi+3 \arg(1-i \sqrt{3} \cos{\theta})$.
The flux penetrating one hexagon is determined as $-2 \phi$.
Note that the net flux through a unit
cell vanishes due to the cancellation of contribution of two triangles
 and a hexagon. It should be also noted that time-reversal symmetry is
broken except for the case of $\phi=0,\pi$.

 From now on we choose the unit of $t \cos \frac{\theta_{ij}}{2}=1$,
 and set the length of each bond as unity. We define three vectors
$\vec{a_1}=(-1/2,-\sqrt{3}/2),\vec{a_2}=(1,0),
 \vec{a_3}=(-1/2,\sqrt{3}/2)$, which  represent the displacements in a
 unit cell from A to B site, from B  to C site, from C 
 to A site, respectively. In this
notation, the Brillouin zone (BZ) is a hexagon with the corners of
$\vec{k}=\pm (2 \pi/3) \vec{a_1}, \pm (2 \pi/3) \vec{a_2},
\pm (2 \pi/3) \vec{a_3}$, two of which are independent.

To diagonalize the Hamiltonian, we rewrite
the Hamiltonian in the momentum space as $H= \sum_{\vec{k}}
\psi^{\dag}(\vec{k})h(\vec{k})\psi(\vec{k})$, where
$\psi(\vec{k})={}^t(\psi_A(\vec{k}),\psi_B(\vec{k}),\psi_C(\vec{k}))$ 
and $h(\vec{k})$ is a $3 \times 3 $ matrix:
\begin{equation}
h(\vec{k})=2 \left(
     \begin{array}{ccc}
         0 &  \cos(\vec{k} \cdot \vec{a_1})/ \omega &
              \cos(\vec{k} \cdot \vec{a_3}) \omega\\
              \cos(\vec{k} \cdot \vec{a_1})\omega& 0 &
                  \cos(\vec{k} \cdot \vec{a_2})/ \omega\\
                  \cos(\vec{k} \cdot \vec{a_3})/ \omega &
                  \cos(\vec{k} \cdot \vec{a_2}) \omega & 0                       
         \end{array}
         \right)   ,            
\end{equation}
where $\omega=e^{i  \phi/3}$
After diagonalization, we obtain eigenvalues $E_i$ and eigenvectors
$\vert \psi_i(\vec{k}) \rangle=(a_i(\vec{k}) \psi_A^{\dag}(\vec{k})+
b_i(\vec{k})
\psi_B^{\dag}(\vec{k})+c_i(\vec{k}) \psi_C^{\dag}(\vec{k})) \vert 0
\rangle$,
which satisfy
$h(\vec{k}) \vert \psi_i(\vec{k}) \rangle=E_i(\vec{k}) \vert \psi_i(\vec{k})
\rangle$.
There are three 
bands with dispersion relations 
\begin{eqnarray}
&E_{{\rm upper}}(\vec{k})&=4 \sqrt{\frac{1+f(\vec{k})}{3}}
\cos\frac{\theta(\vec{k})}{3}, \nonumber \\
&E_{{\rm middle}}(\vec{k})&=4 \sqrt{\frac{1+f(\vec{k})}{3}}
\cos\frac{\theta(\vec{k})-2\pi}{3}, \label{eqn:energy}\\
&E_{{\rm lower}}(\vec{k})&=4 \sqrt{\frac{1+f(\vec{k})}{3}}
\cos\frac{\theta(\vec{k})+ 2 \pi }{3} \nonumber,
\end{eqnarray}
where $\theta(\vec{k}) (0 \le \theta(\vec{k}) \le \pi)$ is defined by
the argument of
\begin{equation}
 f(\vec{k}) \cos\phi+i \sqrt{4
\left( \frac{1+f(\vec{k})}{3} \right) ^3-
(f(\vec{k}) \cos\phi)^2} ,
\end{equation}
and $f(\vec{k})$ is given by
$f(\vec{k})=2 \cos(\vec{k} \cdot\vec{a_1}) \cos(\vec{k} \cdot\vec{a_2})
\cos(\vec{k} \cdot \vec{a_3})$.

\begin{figure}
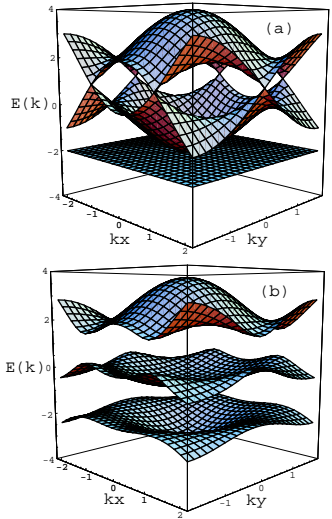

\begin{center}
\vspace{0mm}
\hspace{0mm}
\epsfxsize=4.3cm
\epsfbox{ene0.eps}
\vspace{0mm}
\hspace{0mm}
\epsfxsize=4.3cm
\epsfbox{enepi3.eps}
\caption{The energy spectra eq.(\protect\ref{eqn:energy}) in the case of
(a) $\phi =0$, and (b) $\phi=\pi/3$.}
\label{energy}
\end{center}
\end{figure}

As special cases, the energy dispersion for $\phi=0,\pi/3$
are shown in  Fig.\ref{energy}.
The  spectra have some characteristic features. The relation 
$E_{{\rm lower}}(\vec{k}) \le E_{{\rm middle}}(\vec{k}) \le
E_{{\rm upper}}(\vec{k})  $ is
always satisfied, and
the equality is achieved only when the system is time-reversal
symmetric,
i.e.,$\phi=0,\pi$.
When $\phi=0$, the lower  band becomes dispersionless
 $(E_{{\rm lower}}(\vec{k})=-2=const.)$, which is the reflection of the fact
that the Kagome lattice is a line graph of the honeycomb structure
\cite{mielke}. This flat band 
touches at the center of the BZ $(\vec{k}=0)$ with
the middle band, whose dispersion relation around it is $E_{{\rm middle}}
(\vec{k})
\propto \vec{k}^2 $. The middle band and the upper  band touch at 
two independent corners of the BZ. Around each of the two corners,
the dispersion is
expressed by a massless Dirac fermion.
The spectrum of  $\phi=\pi$ is particle-hole conjugate of that of
$\phi=0$; therefore, the  upper band becomes flat with an eigenvalue of 2.
Generally the energy spectra has no particle-hole symmetry except for the
case of $\phi=\pm \pi/2$, in which the middle band becomes
dispersionless: $E_{{\rm middle}}(\vec{k})=0$.

We now calculate the Hall conductance of this system. It is clear that
the Hall conductance $\sigma_{xy}$ is equal to zero $(\sigma_{xy} =0)$
in the time-reversal symmetric cases $\phi=0,\pi$.  Therefore, we 
focus on the case of $\phi \neq 0,\pi$. In this case  there is an
energy gap between each band, and we   first assume that the  Fermi energy is
lying in the gap. The Hall conductance is given by the
summation of that for each band below the Fermi energy:
$\sigma_{xy}=\sum_{E_i \leq E_F}  \sigma_{xy}^{i}$, and
the Hall conductance is generally quantized, i.e., $\sigma_{xy}=\nu e^2/h$
($\nu$: integer) \cite{tknn}.
The contribution to the Hall conductance from an $i$-th band is written
as
\begin{equation} 
\sigma_{xy}^{i}=\frac{e^2}{h} \frac{1}{2 \pi i}\int_{BZ}d^2 k \ \hat{z}
\cdot \nabla_{\vec{k}} \times\vec{A_i}(\vec{k}) =\frac{e^2}{h}C_i,
\label{eqn:hall}
\end{equation}
where $\vec{A_i}(\vec{k})$ is the vector potential defined with the
$i$-th Bloch wave function as
\begin{equation}
\vec{A_i}(\vec{k})=(a_i(\vec{k}),b_i(\vec{k}),c_i(\vec{k}))^* \cdot
\nabla_{\vec{k}}
{}^t(a_i(\vec{k}),b_i(\vec{k}),c_i(\vec{k})),
\end{equation}
and $C_i$ is the so-called first  Chern number. This value is invariant
under gauge transformation $\vert \psi_i ^{\prime} (\vec{k}) \rangle=
e^{ig(\vec{k})} \vert \psi_i (\vec{k}) \rangle,
\vec{A_{i}^{\prime}}(\vec{k})=\vec{A_{i}}(\vec{k})+i \nabla_{\vec{k}}
g(\vec{k})$,
where $g(\vec{k})$ is
an arbitrary smooth function of $\vec{k}$.
To calculate the Hall conductance explicitly,
we fix the gauge, for example by setting $a_i(\vec{k})$ to be  real.
If this gauge choice is
applicable in the whole region of the BZ, the evaluation
of eq.(\ref{eqn:hall}) leads to $\sigma_{xy}^i=0$. However, generally speaking,
there might be  some points
where  the amplitude of $a_i(\vec{k})$ becomes zero. We call these
points as the center of  vortices. At the center of the vortices, our previous
choice of the gauge is ill-defined, and
we have to choose another gauge, for example $b_i(\vec{k})$ is real
around them.
It is this phase mismatch between patches in the BZ that
contributes to the non-zero Hall conductance.

In our model, we can calculate the Hall conductance  of each band
analytically. We take the lower 
band as an example and we will omit the band index in this paragraph.
We choose the gauge
of real $a(\vec{k}) $, and rewrite the
eigenvector as $(a(\vec{k}),b(\vec{k}),c(\vec{k}))=(a^{\prime}(\vec{k}),
b^{\prime}(\vec{k})e^{-i \xi_{b}(\vec{k})},
c^{\prime}(\vec{k})e^{-i \xi_{c}(\vec{k})})$
, where $a^{\prime}(\vec{k})>0,b^{\prime}(\vec{k})>0,
c^{\prime}(\vec{k})>0, \xi_{b}(\vec{k}), \xi_{c}(\vec{k})$
are real numbers. This gauge
choice is ill-defined at the point $\vec{k}=(0,\pi/\sqrt{3})$;
therefore, we
take the gauge of real $b(\vec{k})  $ around it.  The
first Chern number is written as
\begin{equation}
C=\frac{1}{2 \pi} \oint _{\Gamma} d \vec{k} \cdot \nabla_{\vec{k}}
\xi_{b}(\vec{k}),
\end{equation}
where the integral is over the closed loop $\Gamma$ around  the vortex.
An explicit calculation leads to $C_{{\rm lower}}=-{\rm sgn}(\sin \phi)$.
In a similar way, we can calculate the contribution from the middle and
the upper band, and the results are $C_{{\rm middle}}=0,
C_{{\rm upper}}={\rm sgn}(\sin \phi)$. This means that
quantum Hall effect with zero total flux in the unit cell is realized in
the present model \cite{haldane}.  We have also confirmed these results
numerically.

It is  noted here that an  infinitesimal tilting of spin and hence the
spin chirality $\phi$ opens the gap, and the bands obtain chiralities.
Although this situation is similar to the chiral spin liquid \cite{wwz},
still there are crucial differences between these two cases.
In the present model,  both the spin direction on each site and the flux
through a plaquette are ordered. This is in sharp contrast to the
chiral spin liquid where only the flux through a plaquette is ordered,
and the direction of spin on each site is fluctuating. 
Furthermore, in the present case, the physical observable $\sigma_{xy}$
is nonvanishing and quantized while $\sigma_{xx}$ is zero. In the chiral spin
liquid, on the other hand, $\sigma_{xy}$ and $\sigma_{xx}$  always vanish
because charge degree of freedom
is missing there.  Even when carriers are doped and the anyon superconductivity
occurs \cite{laughlin}, the Meissner term in the
action, i.e., $\rho_{s} \vec{A}
\cdot \vec{A}$, is  dominant and detection of $\sigma_{xy}$ through electric
Hall effect is difficult.

It is proved that the ground 
state of the Hubbard model on the Kagome lattice is a ferromagnet if the flat
band is half-filled \cite{mielke}, and this flat-band  ferromagnetism is
robust under introduction of small dispersion to the dispersionless
band \cite{kusakabe}.
Furthermore the spin-orbit coupling gives the spin anisotropies, which
are different for three crystallographically independent sites.
This introduces the tilting of the spins from the perfect ferromagnetic
alignment, as was assumed in eq.(\ref{eqn:tight}).
Therefore, from the theoretical point of view, we can strongly expect
that once the electron density is $1/3$ per atom on the Kagome
lattice, the chiral spin state presented here is realized and 
Hall conductance is  quantized as  $\sigma_{xy}= \pm e^2/h$.

Finally, we discuss the recent experiments on
$R_2$Mo$_2$O$_7$($R=$Nd, Sm, Gd)  \cite{taguchi,katsufuji},
which are  itinerant ferromagnets on the verge of Mott transition
on the pyrochlore lattice. The spin polarization is almost 
perfect \cite{taguchi}, and the 
tight binding model eq.(3) is the appropriate one for these spin-polarized 
electrons.
These compounds show metallic behaviors, which means that the Fermi energy 
is not in the band gap, 
and the above argument is not straightforwardly applicable to discuss
the large nonvanishing AHE at zero temperature. However,
our results show that each band takes the chiral nature
in the ground state, and we can
expect finite anomalous Hall conductance even when the Fermi energy
is lying inside the band, which is qualitatively consistent with
the experiment. In this case, the magnitude of $\sigma_{xy}$ depends
on band dispersion and also the lifetime of the quasiparticles.
Thus, the quantitative discussion is beyond the scope of the 
present our analysis.  Considerations of these
issues as well as the extension to 3D systems \cite{kohmotohalperin}
are left for future studies. 

In summary, we studied the chiral spin state realized in the flat-band
ferromagnet with spin anisotropy on the 2D Kagome lattice. 
If the Fermi energy is lying in the
gap, we expect quantized Hall conductance $\sigma_{xy}=
\pm e^2/h$ {\it without} external magnetic field.
In  other cases, the system behaves as an
itinerant ferromagnet with finite Hall conductance 
at zero temperature. This feature is qualitatively in
good accordance with recent experiments on the pyrochlore oxides
$R_2$Mo$_2$O$_7(R=$Nd, Sm, Gd$)$.

The authors acknowledge Y.~Tokura,  Y.~Taguchi, and A.~Abanov 
for fruitful discussions. 
This work  is supported by 
 Grant-in-Aid for Scientific Research on Priority Areas 
and Grant-in-Aid for COE research 
from the Ministry of Education, Science, Culture and Sports of Japan,

\end{multicols}
\end{document}